\documentclass{aa501}
\usepackage{graphics,amssymb,natbib,rotating,psfrag,amsmath,epsfig,array}

\title{Temporal properties of the short gamma-ray bursts}
\titlerunning{Short gamma-ray bursts}

\author{S.\,McBreen\inst{1} \and
        F.\,Quilligan\inst{1} \and
        B.\,McBreen\inst{1} \and
        L.\,Hanlon\inst{1} \and
        D.\,Watson\inst{2}}

\institute{Department of Experimental Physics, University College Dublin, Dublin 4, Ireland
    \and X-Ray Astronomy Group, Department of Physics and Astronomy, Leicester University, Leicester LE1 7RH, UK}

\offprints{smcbreen@bermuda.ucd.ie}
\date{Received / Accepted }
\abstract{A temporal analysis has been performed on a sample of
100 bright gamma-ray bursts (GRBs) with $T_{90}$$<$$2\,$s from the
BATSE Current Catalog.  The GRBs were denoised using a median
filter and subjected to an automated pulse selection algorithm as
an objective way of identifing the effects of neighbouring
pulses.  The rise times, fall times, FWHM, pulse amplitudes and
areas were measured and the frequency distributions are presented
here.  All are consistent with lognormal distributions.  The
distribution of time intervals between pulses is not random but
consistent with a lognormal distribution.  The time intervals
between pulses and pulse amplitudes are highly correlated with
each other.  These results are in excellent agreement with a
similar analysis that revealed lognormal distributions for pulse
properties and correlated time intervals between pulses in bright
GRBs with $T_{90}$$>$$2\,$s.  The two sub-classes of GRBs appear
to have the same emission mechanism which is probably caused by
internal shocks.  They may not have the same progenitors because
of the generic nature of the fireball model.
}

\begin{document}
 \maketitle

\section{Introduction}

It has been recognised that GRBs may occur in two sub-classes
based on spectral hardness and duration with $T_{90}$$>$$2\,$s
and $T_{90}$$<$$2\,$s \citep{kmf:1993,dbt:1995,paciesas:2001}. The
bimodal distribution can be fit by two Gaussian distributions in
the logarithmic durations \citep{mhlm:1994}.  There is significant
evidence for a third subgroup as part of the long duration GRBs
\citep{mfb:1998,horv:1998} but this has been questioned because of
a possible BATSE selection effect \citep{hhp:2000}.  It has been
suggested \citep{cmo:1999} that the small group of GRBs with
T$_{90} <$ 0.1 s form an additional category.  The short GRBs
have a higher value of $\langle V/V_{\rm  max}\rangle$ \citep{kc:1996}, a much smaller
value of the spectral lag \citep{nsb:2000}, a pulse shape that
depends on position in the burst \citep{gupta:2000} and a smaller
space density than long GRBs \citep{schmidt:2001}.

A variety of statistical methods have been applied to the
temporal properties of GRBs with $T_{90}$$>$$2\,$s.  It is
important to compare the temporal profiles of the long and short
GRBs to determine the similarities and differences between the
two classes in an objective way.  A detailed objective analysis
has been performed on the temporal profiles of a large sample of
319 bright GRBs with $T_{90}$$>$$2\,$s
\citep{quillig:2001,hmq:1998}.  The properties of the pulses in
GRBs and the time intervals between them were found to be
consistent with lognormal distributions.  These results can be
used as templates for comparison with a similar analysis of GRBs
with $T_{90}$$<$$2\,$s.

The analysis method is presented in section 2 and the results in
section 3.  In section 4 the results are discussed and compared
with the sample of long GRBs.

\begin{table*}[htbp]
 \caption{
        The parameters of the pulse properties in GRBs include the median value for the data,
        the median $\mu$ and the standard deviation $\sigma$
        expressed as natural logarithms, the width of
        the lognormal distributions at the $\pm50\%$ level in normal space, 
	the KS probability that properties of pulses selected at different levels of $\tau_{\rm \sigma}$ and $\tau_{\rm i}$
	were drawn from the same
	distribution as those selected at $\mathrm{\tau}_{\rm \sigma}$ $\geq$ 5, $\mathrm{\tau}_{\rm i}$ $\geq$ 50\%.
        All pulses with $\mathrm{\tau}_{\rm \sigma}$ $\geq$ 5 were used for the
        time intervals and not just isolated pulses. The values
        for the pulse areas and amplitudes were obtained from 55 GRBs that were summed
        over two BATSE Large Area Detectors. 
        }
\setlength{\tabcolsep}{0.3cm}
\begin{tabular}[b]{lccccccc}
\vspace{0.1cm} Property   & Median & $\mu$ & $\sigma$ & Width
($\pm50\%$)  & KS ($\mathrm{\tau}_{\rm \sigma}$ $\geq$ 3,$\mathrm{\tau}_{\rm i}$ $\geq$ 20\%)  &
 KS ($\mathrm{\tau}_{\rm \sigma}$ $\geq$ 8, $\mathrm{\tau}_{\rm i}$ $\geq$ 80\%) &
\\

\hline 
Rise Time & 0.035 & $-$3.31 &  0.94 & 0.012-0.11 & 0.48 & 0.16 &  \\
Fall Time & 0.056 & $-$2.89 & 0.98 &  0.017-0.176 &  0.07 & 0.19 & \\ 
FWHM  & 0.045 & $-$3.17 & 1.01 & 0.013-0.138 & 0.73 & 0.48  & \\ 
Area       & 1.5$\times 10^{5}$& 11.7 & 1.24 & 28-520($\times10^{3}$) & 0.44 & 0.15 & \\ 
Pulse Amplitude & 1.02$\times 10^{4}$ & 9.22 & 0.83 &  3.8-27($\times10^{3}$)  & 0.90 & 0.72 &\\ 
Asymmetry Ratio& 0.65 & $-$0.42  & 0.91 &0.19-2.22  & 0.78   & 0.30 &  \\ 
$\Delta$T  & 0.095 & $-$2.24 & 0.87 & 0.038-0.3   & 0.83 &  0.96  &   \\
 \hline
        \end{tabular}
\end{table*}

\section{Data Preparation}
The dataset was taken from the BATSE current catalog.  The time
tagged event data at 5\,ms resolution was used.  The data from
the four energy channels were combined into a single channel to
maximise the signal to noise ratio.  A subset of the current
catalogue was selected with GRB durations less than two seconds
($T_{90}<2$\,s).  These GRBs were ordered according to their peak
flux in 256\,ms ($P_{\rm 256 ms}$) and the first 100 bursts
without data gaps and with 5\,ms data available for the complete
burst where selected .  All GRBs in the sample had $P_{\rm 256
ms}$ $>$ 1.6 photons cm$^{-2}$s$^{-1}$.

\subsection{Background subtraction}

The first step in the data preparation involved selecting the appropriate background for subtraction.  The start and end times for each
burst were identified .  A background section was then identified which was usually after the main section of the GRB.  If 5\,ms data
was available for the background level estimation, it was used, otherwise a 64\,ms section was selected.  The median value of the data
points in the background section was subtracted from all other values in the active section of the GRB.

\subsection{Denoising and pulse selection}

A median filter was used to denoise the GRBs.  This denoises a
signal by finding the median of each bin.  The best denoised
signal was found by varying the bin size and the value of
$\sigma$.  The wavelet method \citep{quillig:2001} was not used
because the short durations of the GRBs produced a smaller number
of data points than that required for our wavelet method.

The pulse selection method was described elsewhere
\citep{quillig:2001}.  The pulses selected had a threshold of 5
$\sigma$ ($\mathrm{\tau}_{\rm \sigma}$ $\geq$ 5) and isolated from
adjacent pulses by at least 50\% ($\mathrm{\tau}_{\rm i}$ $\geq$
50\%).  A value of $\mathrm{\tau}_{\rm i}$ $\geq$ 50\% implies that the
two minima on either side of the pulse maximum must be at or
below half the maximum value.  A total of 313 pulses were
selected with $\mathrm{\tau}_{\rm \sigma}$ $\geq$ 5 and 181 of these
had $\mathrm{\tau}_{\rm i}$ $\geq$ 50\%.

\section{Results}

\subsection{Distributions of $t_{\rm r}$, $t_{\rm f}$, FWHM, pulse area, pulse amplitude and $\Delta$T}
The distribution of the number of pulses with
$\mathrm{\tau}_{\rm \sigma}$ $\geq$ 5 is given in Fig. 1.  The median
value of the number of pulses per GRB is 2.5 and is smaller than
the value of 6 for long GRBs \citep{quillig:2001}. The
distributions of rise time $t_{\rm r}$, fall time $t_{\rm f}$ and full width
at half maximum (FWHM) for the isolated pulses are presented in
Fig. 2 along with the lognormal fits to the data. The
distributions of pulse amplitudes and areas are presented in
Fig. 3 for the isolated pulses.  The distribution of time
intervals between pulses with $\mathrm{\tau}_{\rm \sigma}$ $\geq$ 5 is
given in Fig. 4.

\begin{figure}[hbp]
    \leavevmode
    \psfrag{xlab}[b]{\small Number of Pulses}
    \psfrag{ylab}[b]{\small Number of GRBs}
    \vspace{2em}
\includegraphics[width=0.85\columnwidth]{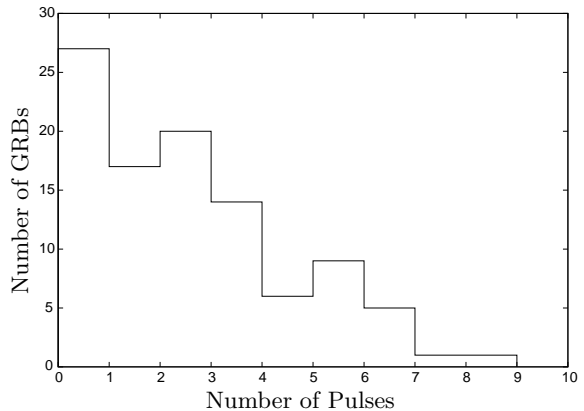}
\label{fig:grbs_pulses}

\vspace{.5em}
    \caption{
        The number of pulses versus number of $\gamma$-ray bursts.
      }
\end{figure}

The median values and the parameters of the lognormal fits are
presented in Table 1.  The distribution of the pulse asymmetry
ratios, $t_{\rm r}$/$t_{\rm f}$ , is not presented but the values are listed
in Table 1.  The pulse fall time is generally longer than the
rise time.  The widths or ranges of the lognormal distributions at the $\pm50\%$ 
are are also given in Table 1.
The Kolmogorov-Smirnov (KS) test was used to indicate
the effects of varying $\mathrm{\tau}_{\rm \sigma}$ and 
$\tau_{\rm i}$. The pulse property distributions at $\mathrm{\tau}_{\rm \sigma}$ $\geq$ 5
and $\tau_{\rm i}$ $\geq$ 50\% were compared with the distributions 
selected over large range of $\mathrm{\tau}_{\rm \sigma}$ and $\tau_{\rm i}$.
The significance level probabilities from the KS test for the comparison are presented
in Table 1 for $\mathrm{\tau}_{\rm \sigma}$ $\geq$ 3, $\tau_{\rm i}$ $\geq$ 20\% and 
$\mathrm{\tau}_{\rm \sigma}$ $\geq$ 8, $\tau_{\rm i}$ $\geq$ 80\% and all have acceptable values.
It was found that when $\tau_{\rm i}$ increased 
the pulses become less contaminated by overlaps
from adjacent pulses. 

\begin{figure}[htbp]
    \psfrag{xlab}[b]{\small $t_{\rm r}$ (sec)}
    \psfrag{ylab}[b]{\small frequency}
    \vspace{1.5em}
\includegraphics[width=0.85\columnwidth]{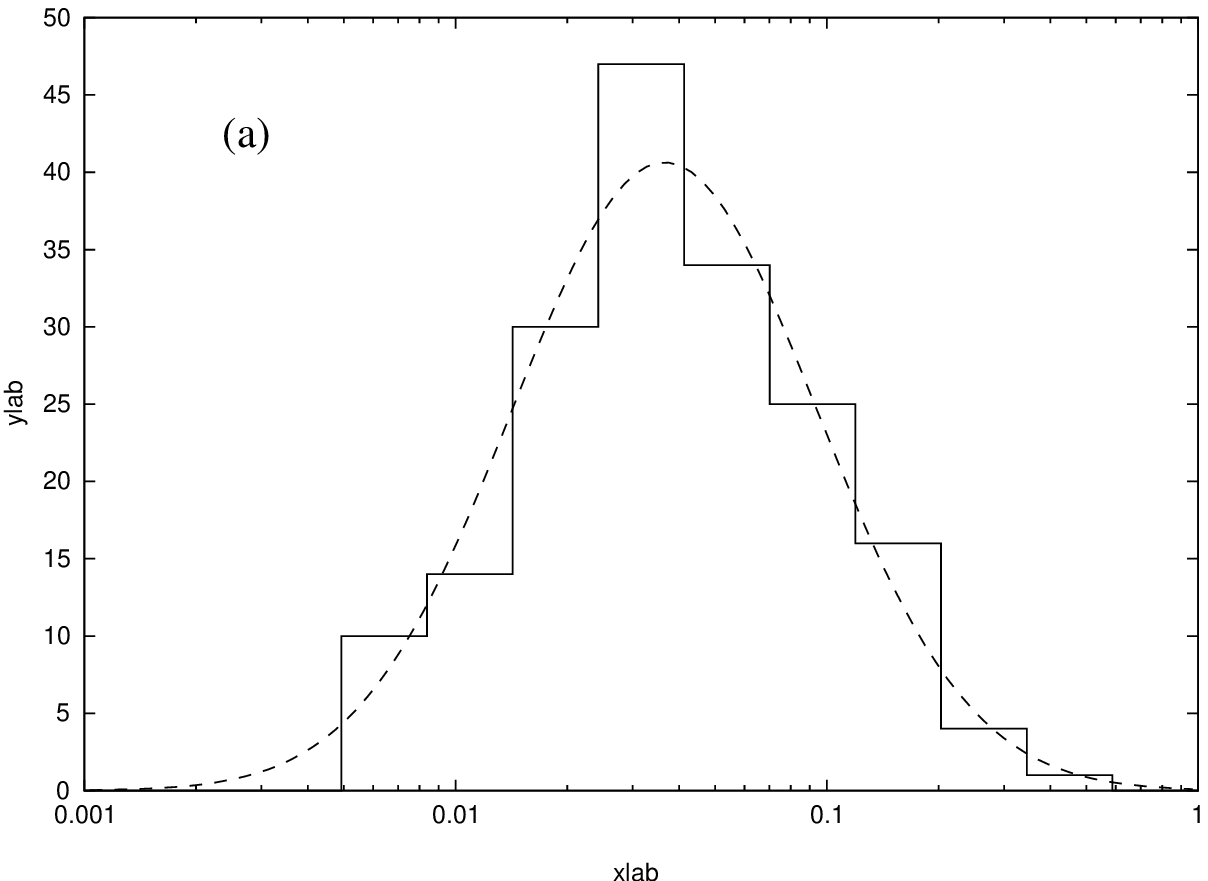}

    \psfrag{xlab}[b]{\small $t_{\rm f}$ (sec)}
    \psfrag{ylab}[b]{\small frequency}
\includegraphics[width=0.85\columnwidth]{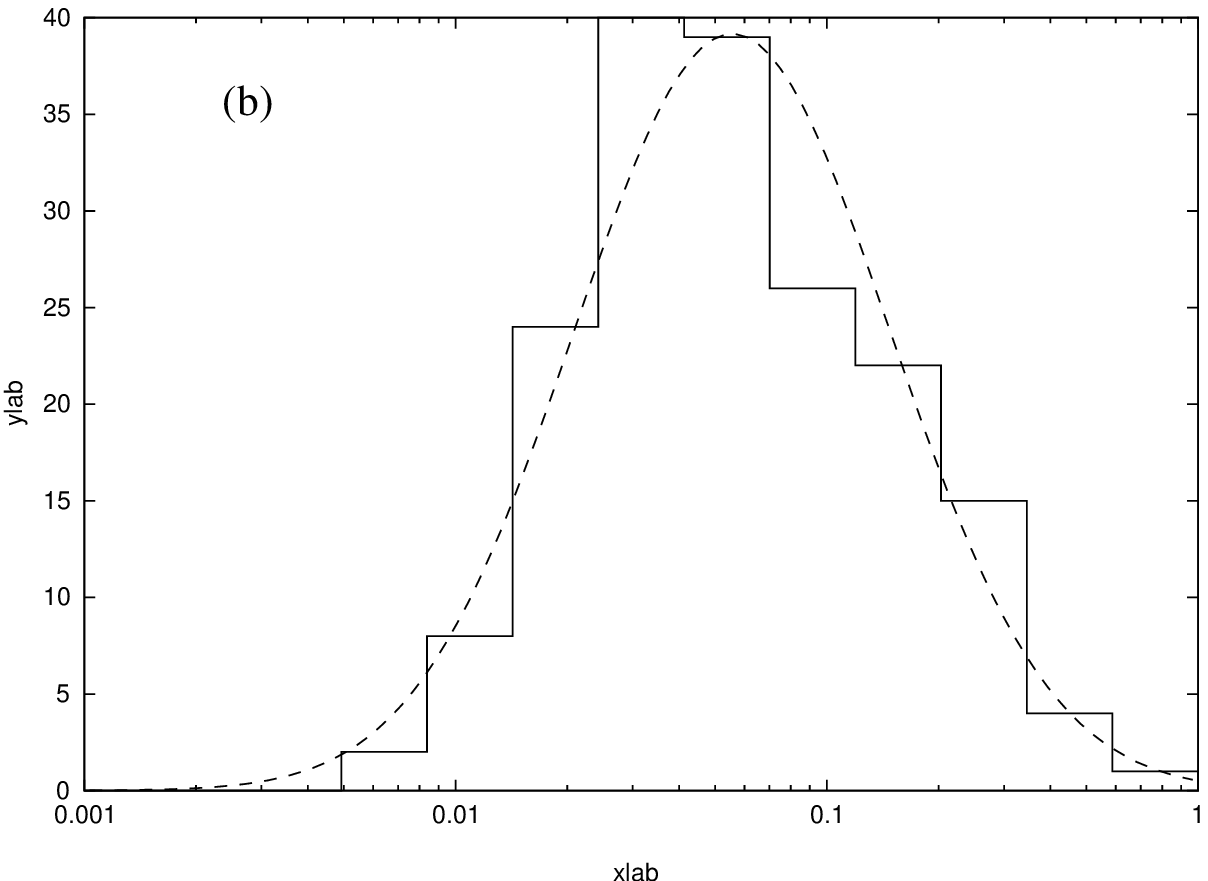}

    \psfrag{xlab}[b]{\small FWHM (sec)}
    \psfrag{ylab}[b]{\small frequency}
\includegraphics[width=0.85\columnwidth]{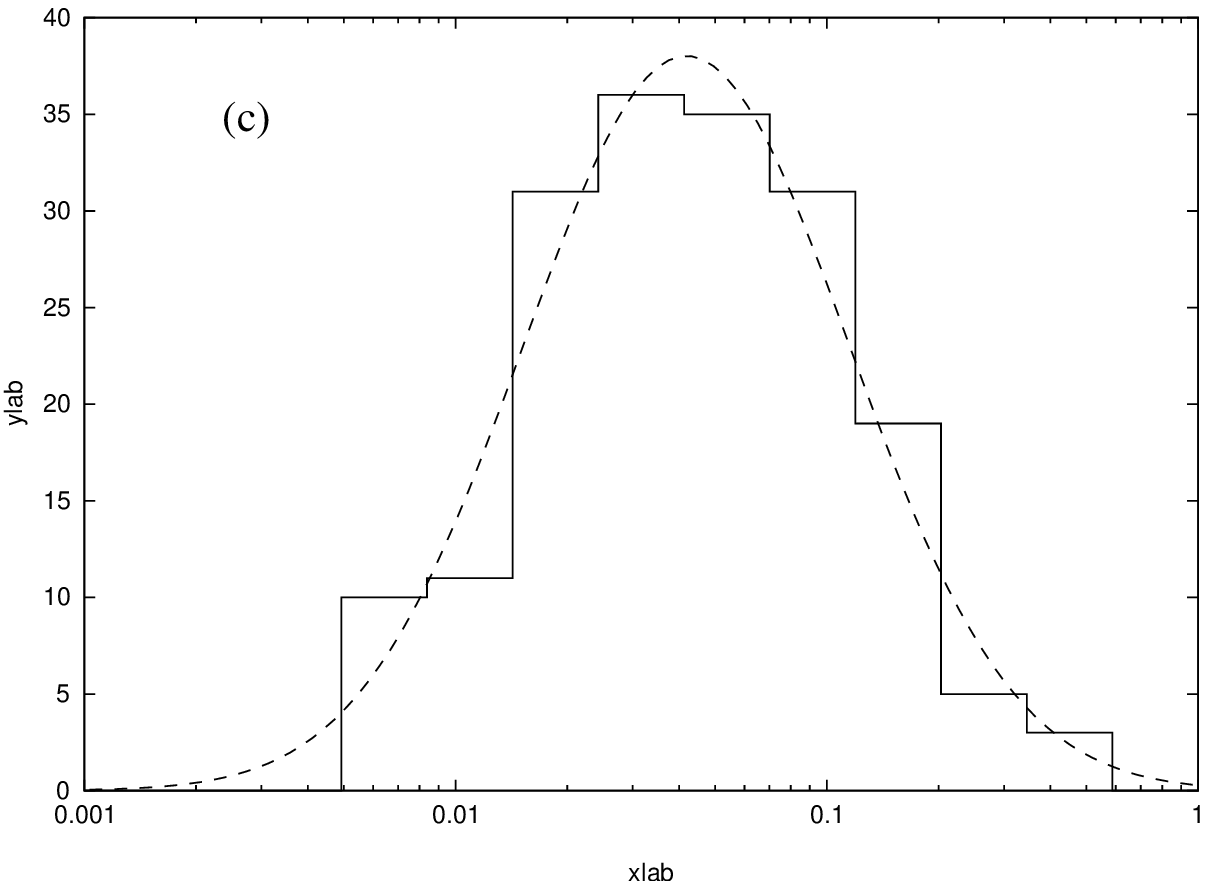}

    \caption{%
        The distribution of $t_{\rm r}$, $t_{\rm f}$, and FWHM ({a-c}) for pulses with $\mathrm{\tau}_{\rm \sigma}$ $\geq$ 5
        and  $\mathrm{\tau}_{\rm i}$ $\geq$ 50\%. The dashed lines are lognormals that are consistent with the data.
       }
\end{figure}

\vspace{1.5em}
\subsection{Correlation Analysis}

The correlation coefficients between a range of GRB parameters and pulse parameters were obtained using the Spearman Rank Order
correlation coefficents $\rho$ with associated probabilities and the values are listed in Table 2.  The correlation coefficients for
the time intervals between pulses are listed in Table 3 and they are strongly correlated with each other.

\begin{table}[h]
 \caption{Spearman Rank Order correlation coefficients $\rho$ and associated
 probabilities between a range of burst (first 3 rows) and pulse parameters (second 10 rows).}
\setlength{\tabcolsep}{.10cm}
\begin{tabular}{lccc}  \\
Properties & $\rho$ & Probability\\ \hline
No. of Pulses vs. $T_{90}$  & 0.18 & 0.073 & \\
No. of Pulses vs. Total Fluence  & 0.38 & $9.8 \times 10^{-5}$ &
\\ $T_{90}$ vs Total Fluence & 0.21 & 0.037 & \\ \hline
Rise Time vs. Fall Time & 0.54 & $2.7 \times 10^{-15}$ & \\
Rise Time vs. FWHM  & 0.79 & $1.3 \times 10^{-40}$ & \\
Rise Time vs. Area & 0.59 & $3.3 \times 10^{-10}$ & \\
Rise Time vs. Pulse Amplitude  & 0.013 & 0.90 & \\
Fall Time vs. FWHM & 0.75 & $2.6 \times 10^{-33}$ & \\
Fall Time vs. Area & 0.64 & $2.0 \times 10^{-12}$ &
\\ Fall Time vs. Pulse Amplitude  & 0.078 &  0.45 & \\
FWHM vs. Area & 0.60 & $6.89 \times 10^{-11}$ & \\
FWHM vs. Pulse Amplitude & $-$0.05 & 0.64 & \\
Area vs. Pulse Amplitude  & 0.65 & $1.1 \times 10^{-12}$ & \\
\hline
\end{tabular}
\end{table}

\begin{table}[htb]
 \caption{
The Spearman correlation coefficients $\rho$ and probabilities for $\Delta$Ts
(first 2 rows) and PAs
(second 2 rows). The first and third rows refer to adjacent $\Delta$Ts and PAs
while the second and fourth rows refer to $\Delta$Ts or PAs separated by one time interval or pulse.
The first / second entry for the $\Delta$Ts gives $\rho$ when the $\Delta$Ts are unnormalised / normalised
by $T_{90}$. The first / second entry for the PAs gives $\rho$ when the PAs are
unnormalised / normalised by the maximum peak in the burst.
}
\setlength{\extrarowheight}{1pt}
\addtolength{\tabcolsep}{-1pt}
\begin{tabular}{lccccc}  \\
Total Number & $\rho$ & Probability\\
\hline 
140   & 0.42/0.30  & $1.7 \times 10^{-12}$/$3.0 \times 10^{-3}$ & \\ 
84 	&0.48/0.32   & $4.9 \times 10^{-6}$/$3.2 \times 10^{-3}$ & \\ 
\hline 
213/107   & 	0.39/0.24 	& $5 \times 10^{-9}$/$1.1 \times 10^{-2}$ & \\ 
140/84    &	0.13/$-$0.09 	& 0.13/0.42 & \\ 
\hline 
\end{tabular}
\end{table}

\begin{figure}[htbp]
    \psfrag{xlab}[t]{\small Area (counts)}
    \psfrag{ylab}[b]{\small frequency}
    \vspace{1.5em}
\includegraphics[width=0.85\columnwidth]{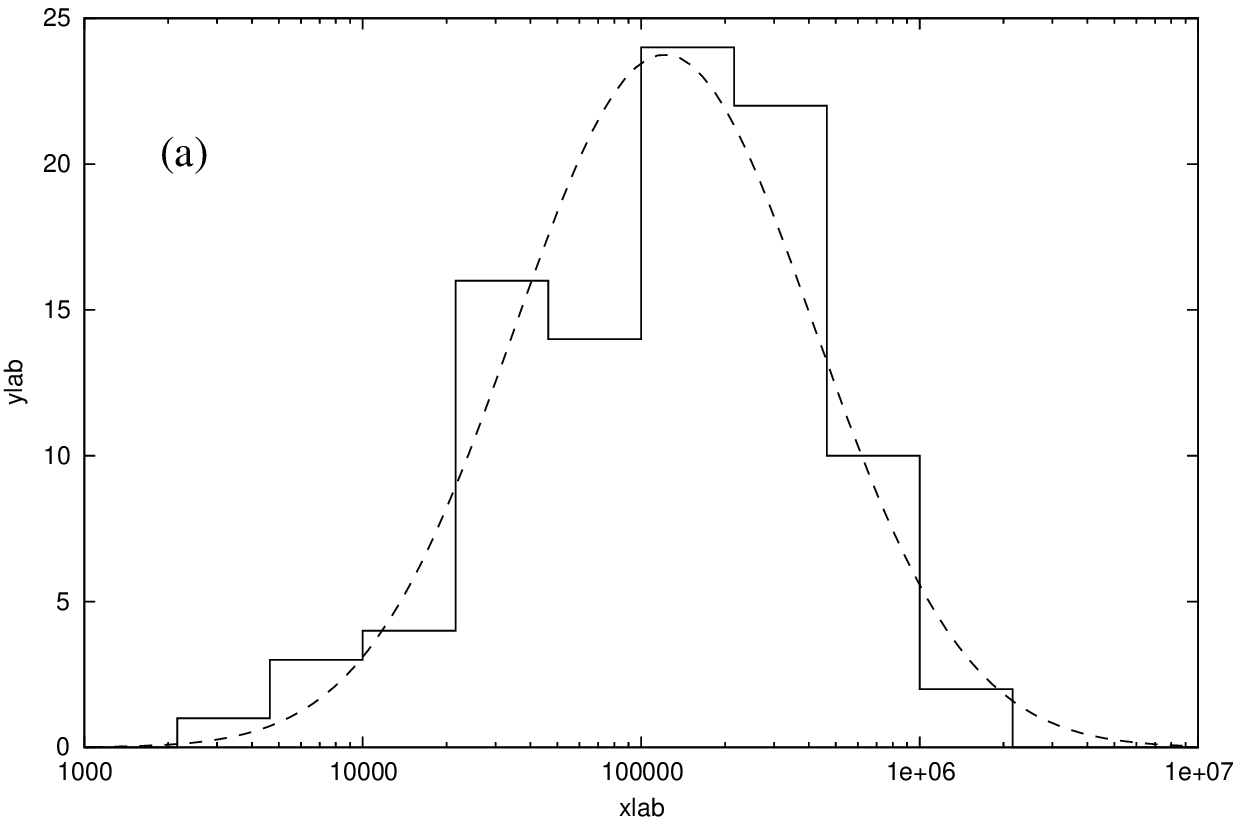}

    \psfrag{xlab}[t]{\small Pulse Amplitude (counts)}
    \psfrag{ylab}[b]{\small frequency}
    \vspace{1.5em}
\includegraphics[width=0.85\columnwidth]{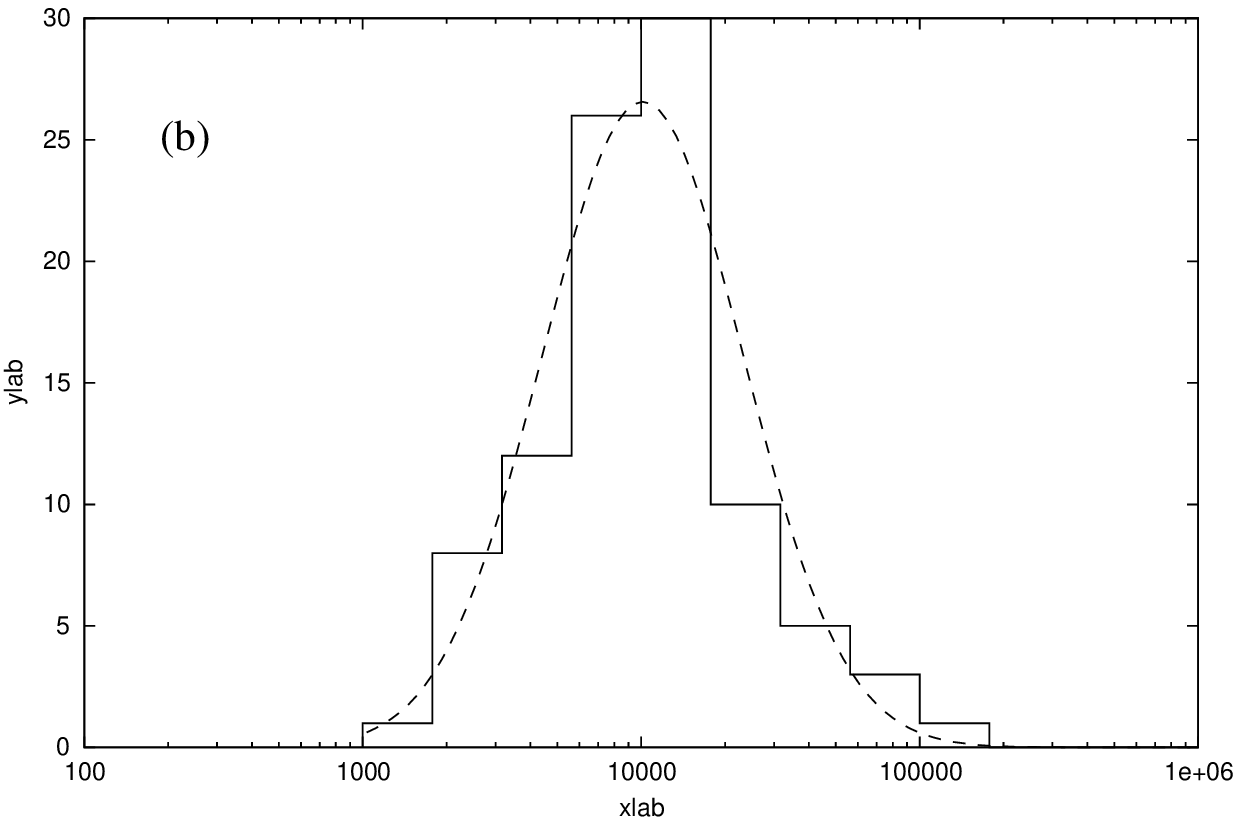}
\vspace{1.5em}
    \caption{
      The distributions of pulse area and amplitude selected at $\mathrm{\tau}_{\rm \sigma}$ $\geq$ 5 and $\mathrm{\tau}_{\rm i}$ $\geq$
      50\% and consistent lognormals (dashed lines). These distributions were generated from 55 GRBs that were
      summed over two BATSE  Large Area Detectors. } 
\end{figure}

\begin{figure}[htbp]
    \leavevmode
    \psfrag{xlab}[t]{\small $\Delta$T (sec)}
    \psfrag{ylab}[b]{\small frequency}

\includegraphics[width=0.85\columnwidth]{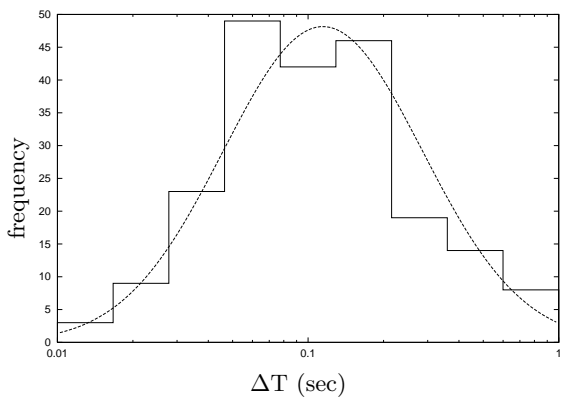}
\label{fig:time_int} \vspace{.5em}
    \caption{
      The distribution of time intervals between the pulses, $\Delta$T,
      selected at $\mathrm{\tau}_{\rm \sigma}$ $\geq$ 5 and consistent lognormal (dashed line).}
\end{figure}

\section{Discussion}

The major result is that pulses in GRBs with $T_{90}$$<$$2\,$s
have very similar distributions of $t_{\rm r}$, $t_{\rm f}$, FWHM, pulse
amplitude, pulse area and $\Delta$T.  The frequency distributions
are broad and all are compatible with lognormal distributions,
and at the 50\% level span an order of magnitude in the
particular pulse property (Table 1).  The same pulse properties
in GRBs with $T_{90}$$>$$2\,$s are also well described by
lognormal distributions \citep{quillig:2001}.  The results of the
correlation analysis on the GRBs and the properties of the pulses
(Table 2) are generally in good agreement between long and short
GRBS.  The main difference is the pulse amplitude because it is
not anticorrelated with $t_{\rm r}$, $t_{\rm f}$ and FWHM.  In GRBs with
$T_{90}$$>$$2\,$s the values of $\Delta$T are not random but
consistent with a lognormal distribution with a Pareto-Levy tail
for a small number of long time intervals in excess of 15\,s. It
was also found that the values of $\Delta$T were correlated over
most of the GRBs and up to more than 20 pulses in the GRBs with
large numbers of pulses.  In short GRBs there is a strong
correlation between the values of $\Delta$T and also between the
pulse amplitudes (Table 3) but the latter are better correlated
in GRBs with $T_{90}$$>$$2\,$s.  There are remarkable
similarities between the statistical properties of the two
classes of GRBs.  The clear conclusion is that the same emission
mechanism accounts for the two types of GRBs.  This conclusion is
in agreement with a very different analysis of the temporal
structure of short GRBs \citep{np:2001}.  The external shock model
\citep{derm:1999} has serious difficulties in accounting for GRBs
with $T_{90}$$<$$2\,$s and with the non-random distribution of
correlated time intervals between pulses.  The results presented
here provide considerable support for the internal shock model
\citep{reemes:1994}.  The internal shock model can account for the
results obtained on long and short GRBs provided the cause of the
pulses and the correlated values of $\Delta$T can be attributed
to the central engine.  It has been shown that the time intervals
between glitches in pulsars and outbursts in SGRs are lognormally
distributed \citep{quillig:2001}.  The similar distributions of
$\Delta$T in short and long GRBs are indirect evidence for
rotation powered systems with super-strong magnectic fields.

There is strong evidence that long GRBs originate in star forming
galaxies providing evidence that they are linked to massive stars
and supernovae \citep{mes:2001}.  A variety of models, such as
collapsars and hypernovae have been proposed as progenitors
(e.g.  MacFadyen \& Woosley 1999\nocite{macfad:1999}).  Models of
these systems imply that jets can be generated to produce GRBs by
an internal shock model. These models cannot account for GRBs
with $T_{90}$ less than about 5\,s.  The shorter GRBs have been
attributed to neutron star (NS) - NS mergers or NS - black hole
mergers (e.g.  Ruffert \& Janka 1999\nocite{ruffjan:1999}).  A
common ingredient of most GRB models is the formation of a black
hole with a temporary torus that accretes and powers the
relativistic jets and the GRBs.  The overall energetics of the
various progenitor models differ by at most an order of
magnitude.  The GRBs appear to have the same emission mechanism
and possibly different progenitors for long and short bursts.

\section{Conclusions}

A sample of bright GRBs with $T_{90}$$<$$2\,$s have been denoised
and analysed by an automatic pulse selection algorithm. The
results show that the distribution of the properties of isolated
pulses and time intervals between all pulses are compatible with
lognormal distributions.  The same mechanism seems to be
responsible for both long and short GRBs and is attributed to the
internal shock model.
\bibliography{katmonic,Dh291.bib}
\bibliographystyle{apj}
\end{document}